# Dielectric and polarization experiments in high loss dielectrics: a word of caution

M. Maglione

ICMCB-CNRS, Université de Bordeaux 1, 87, Av Dr Schweitzer 33806 Pessac France

maglione@icmcb-bordeaux.cnrs.fr

M. A. Subramanian

Department of Chemistry and Oregon State University Materials Institute

Oregon State University Corvallis, OR 97331USA

The recent quest for improved functional materials like high permittivity dielectrics and/or multiferroics has triggered an intense wave of research. Many materials have been checked for their dielectric permittivity or their polarization state. In this report, we call for caution when samples are simultaneously displaying insulating behavior and defect-related conductivity. Many oxides containing mixed valent cations or oxygen vacancies fall in this category. In such cases, most of standard experiments may result in effective high dielectric permittivity which cannot be related to ferroelectric polarization. Here we list few examples of possible discrepancies between measured parameters and their expected microscopic origin.

Among functional materials, oxides are heavily investigated because their wide range of properties. $ABO_3$ perovskites are such a class of materials showing magnetic, ferroelectric, semiconducting, metallic and superconductive ground state on appropriates selection of the A and B cations. Even more interesting, a co-existence between several of these states may be achieved in a given compound. For example, in $BiFeO_3$, Bismuth polarisability and Iron magnetism result in the coexistence of ferroelectricity and anti-ferromagnetism at room temperature which make this material a very promising multiferroic compound [1,2]. In many titanates (B=Ti), strong dielectric permittivity and net lattice polarization are evidenced making these the most studied ferroelectrics [3]. Coexistence between polarisability and conductivity occurs when polyvalent cations such as Fe, Mn and Cu are included in perovskite or in other oxides. Indeed, electron transfer between mixed valent states, for example $Fe^{3+}$ and $Fe^{2+}$, may lead to increased conductivity depending on the oxygen stochiometries and also related to thermal history of the samples such as controlled annealing. Similarly, heterovalent substitution like $Nb^{5+}$ on the $Ti^{4+}$ site profoundly increases the conductivity of $BaTiO_3$ which in fact becomes semi-conducting [4] and even superconducting as in the case of $SrTiO_3$ [5]. In all these cases – multivalent cations, heterovalent substitutions and oxygen vacancies- the materials will exhibit effective dielectric and polarization properties which may not be related to ferroelectricity. In recent years, many materials were shown to display very large or colossal effective dielectric permittivities [6-9]. In this paper, we show that dielectric and polarization probing may be misleading when conductivity and polarisability do co-exist in the same material. Our listing will be split into thermal runs and then isothermal scans.

The standard way of testing a ferroelectric material is to measure induced pyroelectric current when the sample is heated within the ferroelectric state and the observed value is directly related to the ferroelectric polarization. This is why the search for new ferroelectrics usually done by recording pyroelectric current after the sample is poled under electric field cooled conditions. In several cases, a maximum of pyroelectric current and change of it sign on reversing the sign of the poling bias has been taken as an evidence for ferroelectricity [7,10]. There are however many examples in the literature of non-ferroelectric materials which show clearly such a sign reversal of the pyroelectric depolarization currents. Electrets, which include polymer materials [11], are not ferroelectrics and do show pyroelectric current maxima whose sign is following that of the poling field. Such a change of the polarization state resulting from a change of the poling bias is ascribed to the localization of free charges at interfaces in electrets, the so-called space charge effect. Among inorganic oxides, one can find many examples of reversed pyroelectric currents. In figure 1 we show such a reversal experiment reported for $Ta_2O_5$ [12,13], a non ferroelectric oxide widely used in microelectronics. The maximum depolarization current was ascribed to the electron de-trapping from oxygen vacancies. The non-ferroelectric experiment reported in figure 1 compares well with similar results reported for $LuFe_2O_4$, CuO or $CaCu_3Ti_4O_{12}$. In all these cases, heterovalent cations (Fe, Cu) and/or charged oxygen vacancies may act similarly as oxygen vacancies in $Ta_2O_5$ to store electronic charges during the bias cooling and releasing them on recording the current during the heating. In $Ta_2O_5$, $LuFe_2O_4$, CuO or $CaCu_3Ti_4O_{12}$, the change of sign of the net depolarization currents result from a sign change of the global space charge induced by the poling bias : the density of trapped electrons is not homogeneous in the materials having gradients of plus or minus sign and so do the depolarization currents. Such release of space charges is well known in the field of electrets as Thermally Depolarization Currents [11]. In some instances [7, 10], a full detrapping of space charges is achieved at a given temperature which could be mixed with a possible ferroelectric transition. This temperature is however the one where all free charges gets enough thermal energy for the space charge to be cancelled. In other instances, continuous increase of dc current is recorded meaning that semi-conduction starts to be active before a hypothetical depolarization is reached [14]. We probed the depolarization current using a Keithley 610C electrometer and the resistance with a Keithley 195 multimeter. For both experiments, the $CaCu_3Ti_4O_{12}$ ceramics coated with gold electrodes was set in a homemade cell allowing temperature to be scan from 77K up to 800K. For all the reported data, the experimental errors are within the size of the used symbols whenever the current level exceeds $10^{-8}$A. As shown in figure 2, Arrhenius increase of the current is observed which is related to the electrical conduction and not due to depolarization of the ceramics. This is easily confirmed on comparing the current recorded during a heating run after poling (figure 2a) to the dc electrical resistance (figure 2b). Since these two independently measured parameters have exactly the same activation energy, 0.7eV, the former can hardly be called as 'depolarization current'.

While depolarization currents are probing the kinetics of the bias induced space charges, low ac voltage dielectric experiments can be used to investigate the dynamics of localization of space charges. The collective motion of electrical charge induces mesoscopic as well as macroscopic dipoles whose relaxation can be seen on scanning the temperature and sweeping the operating frequency. This is well known in electrets based on polymers [15] and inorganic oxides [16]. In the case of oxides such as $LuFe_2O_4$, one can note that the temperature at which the depolarization currents are maximum is the same as the one where the effective giant dielectric permittivity undergoes a strong dispersion. This relaxation of Debye type nature and thermally activated conductivity which follows the Arrhenius law was reported in all "giant permittivity materials" [6-9]. This similarity indicates a common origin of the giant dielectric permittivity and of its relaxation, the space charges [17]. In the case of $CaCu_3Ti_4O_{12}$, the interplay between grain conductivity and grain boundary insulation was successfully rationalized within the framework of an Internal Grain Boundary Layer model [18] which clearly rejected ferroelectricity as a source of the effective giant permittivity in this compound. In ferroelectrics, however, both the polarization and dielectric permittivity have different behavior. First, it is not only the sense of poling which drives the polarization but also the sign of the temperature slope during the depolarization run [3]. Temperature oscillations in the ferroelectric phase trigger in phase polarization oscillations which modulate the pyroelectric current. This is because the polarization arises from an internally built phase transition and not from the poling step. Also because of this phase transition, the dielectric permittivity never follows Arrhenius type dispersion which is always observed in giant permittivity materials. To summarize this first part, neither the depolarization currents nor the dielectric relaxation are ultimate proofs of ferroelectricity in $Ta_2O_5$, $LuFe_2O_4$, CuO or $CaCu_3Ti_4O_{12}$.

We now turn to isothermal polarization and dielectric experiments which were also used to probe the expected ferroelectric state of giant permittivity materials and in other lossy dielectrics. At a fixed temperature, the polarization hysteresis loop under electric field cycling is the decisive proof of ferroelectricity. Well saturated and squared loops in ferroelectric perovskites can be found in many text books [3]. Since the electric field used to reverse the ferroelectric polarization is rather high, all the materials non-linearities are integrated in a hysteresis loop experiment. One of these non-linearities can again result from the coexistence of conductivity and polarisability in the same compound. In such a case, non-ohmic dc current will occur. When the integrated current is then plotted as a function of electric field, distortions and slight opening of the loop will occur [14]. This should not be confused with the usual coercivity and saturation of ferroelectric hysteresis loops. To clarify this, the whole process of recording hysteresis loops is detailed in figure 3 in the case of the non-ferroelectric $CaCu_3Ti_4O_{12}$ ceramics. To evidence the necessary steps required to remove these extrinsic contributions, we have used a fully analog Sawyer Tower set up including many integration capacitors and resistors to record both the charge and current loops. Moreover, compensating capacitors and resistors brought in parallel to the sample could be tuned manually as to remove spurious impedance contributions to the loops. Details about this experiment can be found in many text books [3]. In figure 3a, the Sawyer Tower integration circuit was used to record charge/electric field loop in this ceramics. At a first sight, this looks like a ferroelectric hysteresis with saturation at about $0.03\mu C/cm^2$ and apparent coercive field of 4kV/cm. To reject dc conductivity from this experiment, one should check the current/electric field loop before integration. This is shown in figure 3b where a strong dc current is seen at 0-electric field. The external circuit is then used to compensate for this dc current then leading to the reduced current loop in figure 3c which has no more contribution at zero bias but which undergoes strong increase at high bias with no sign of bending down. The last step is to turn to the integrated loop in figure 3d which no more looks like a ferroelectric loop because it is nothing but the integration of the semi-conductor shape of the compensated current loop. Whatever the external compensation, it was never possible to get ferroelectric-like loops in the current and charge modes. Since this equivalence is the prerequisite for true ferroelectric behavior, we then firmly conclude that polarization/electric field loops are not due to ferroelectric state in $CaCu_3Ti_4O_{12}$. When true ferroelectric materials such as PZT, $BaTiO_3$, are subjected to the same process (figures 3a to 3d), compensated current and charge loop do lead to consistent ferroelectric polarization reversal. In most of the giant permittivity materials and lossy dielectrics, one should apply the same process in order to reject conductivity contributions to the distortions and opening of electric field loops. Following recent reports by Scott [19], the compensation experiment in figure 3 is one way to distinguish between true ferroelectricity and conductivity induced current loops even when apparently saturated loops are recorded.

Such conductivity contributions can also be seen during isothermal dielectric experiments. In the case of $CaCu_3Ti_4O_{12}$, the real $\varepsilon_1$ and imaginary $\varepsilon_2$ parts of the dielectric permittivity are plotted in figure 4 at room temperature as a function of frequency. On the high frequency side, the bending down of $\varepsilon_1$ and the maximum of $\varepsilon_2$ are the dielectric relaxation which will slow down on cooling. On the low frequency side, the hyperbolic increase of $\varepsilon_2$ stems from the dc contribution. A full fitting of these both contributions using the generalized Jonscher law [20] is shown as continuous curve on both plots. This accurate fitting is a confirmation that charge localization is the origin of the giant effective dielectric permittivity involving electrical conduction as evidenced by impedance plots and can be generalized to many giant permittivity materials [17, 18]. The purpose of this investigation is to indicate that room temperature dielectric spectroscopy is an efficient tool for probing the co-existence of polarisability and conductivity in these materials thus supporting the isothermal charge loops as originating from uncompensated dc currents, not from lattice polarization reversal.

We thus have shown that thermal and isothermal probing of polarization and dielectric permittivity in oxides is not straightforward when residual conductivity is present. Multivalent cations (Fe, Cu, Mn) and/or oxygen vacancies are the sources of this residual conductivity which means that annealing treatments during the processing of ceramics may alter drastically the conductivity and thus the polarization and dielectric parameters. Using well established standard procedures for characterizing ferroelectric materials, one can clearly distinguish between true bulk polarization and extrinsic conductivity contributions.

Acknowledgements Funding from the European commission through the FAME Network Of Excellence and the STREP MACOMUFI is gratefully acknowledged. The research work done at Oregon State University is supported by a grant from National Science Foundation (DMR-0804167). J.P.Chaminade is gratefully acknowledged for providing the Tantalum Oxide ceramics.

Figure captions

Figure 1: zero bias depolarization current of a Tantalum oxide non ferroelectric ceramic. The same depolarization can be induced by UV illumination (see reference 12) or by poling bias during cooling (see reference 13), the sign of the maximum current being changed on reversing the sign of the poling bias.

Figure 2 : (a) depolarization current and (b) resistance of a $CaCu_3Ti_4O_{12}$ ceramic while heating after a poling step. On both plots, lines are fit to an Arrhenius thermal activation. Both activation energies are 0.7eV thus showing that depolarization currents stem from bulk semi-conduction in $CaCu_3Ti_4O_{12}$

Figure 3 (a) apparently saturated charge/electric field loop in a $CaCu_3Ti_4O_{12}$ ceramic at room temperature. (b) current loop corresponding to the previous charge loop; this shows that a strong current flows through the ceramic at vanishing electric field (c) after appropriate compensation from the external circuit, the zero field current has been minimized, the bending of the current at high field stems from non linear conduction (d) after the compensation step of (c), the charge loop does no more look like a saturated hysteresis loop.

Figure 4 real (left scale) and imaginary (right scale) part of the dielectric permittivity of $CaCu_3Ti_4O_{12}$ plotted as a function of frequency at 340K. The low frequency divergence of the imaginary part results from the dc conductivity while the high frequency relaxation is to be related to the grain boundary layer model of ref 18. The line is a fit using a generalized Jonscher model including conductivity and relaxation contributions as described in reference 20.

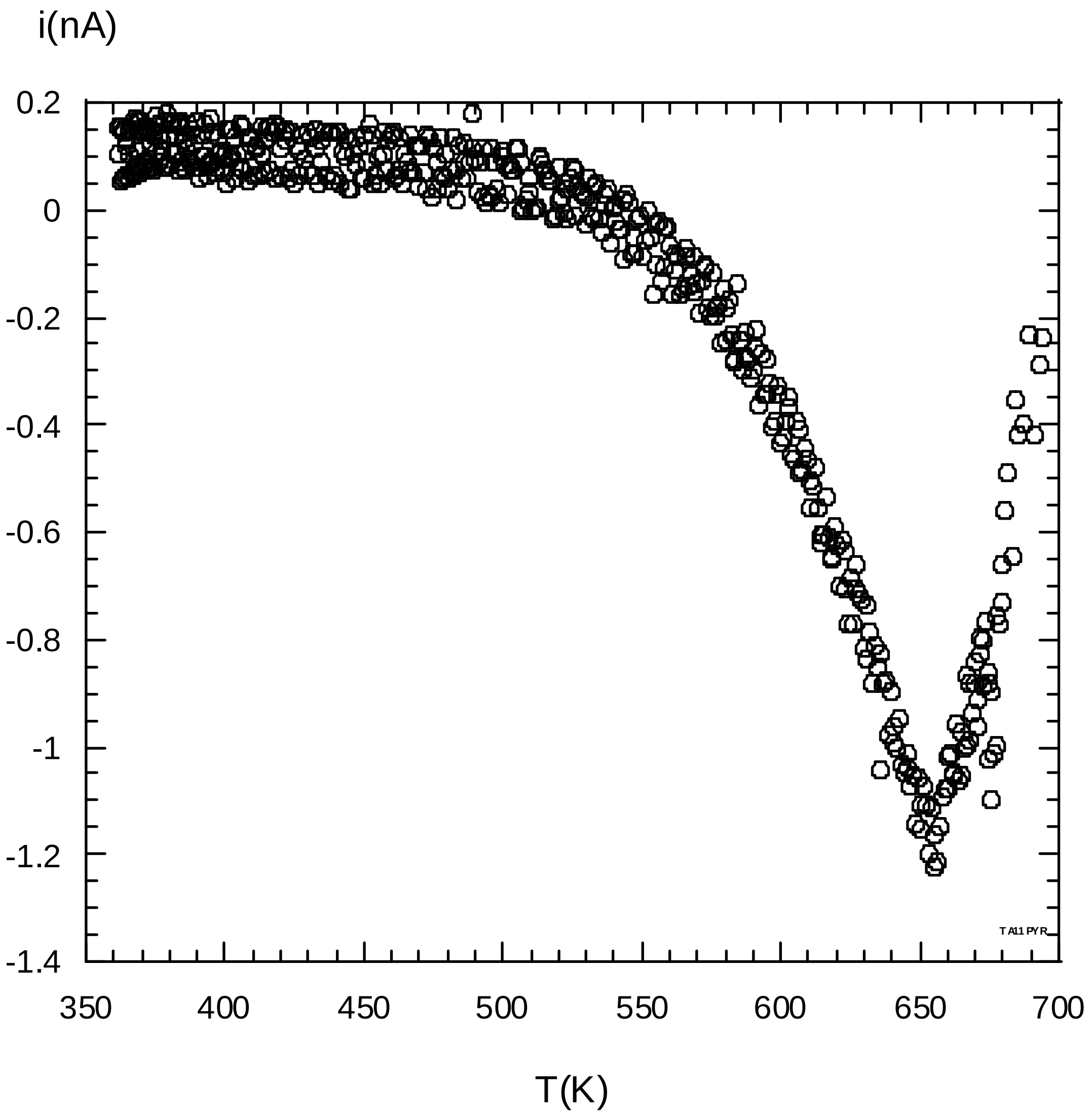

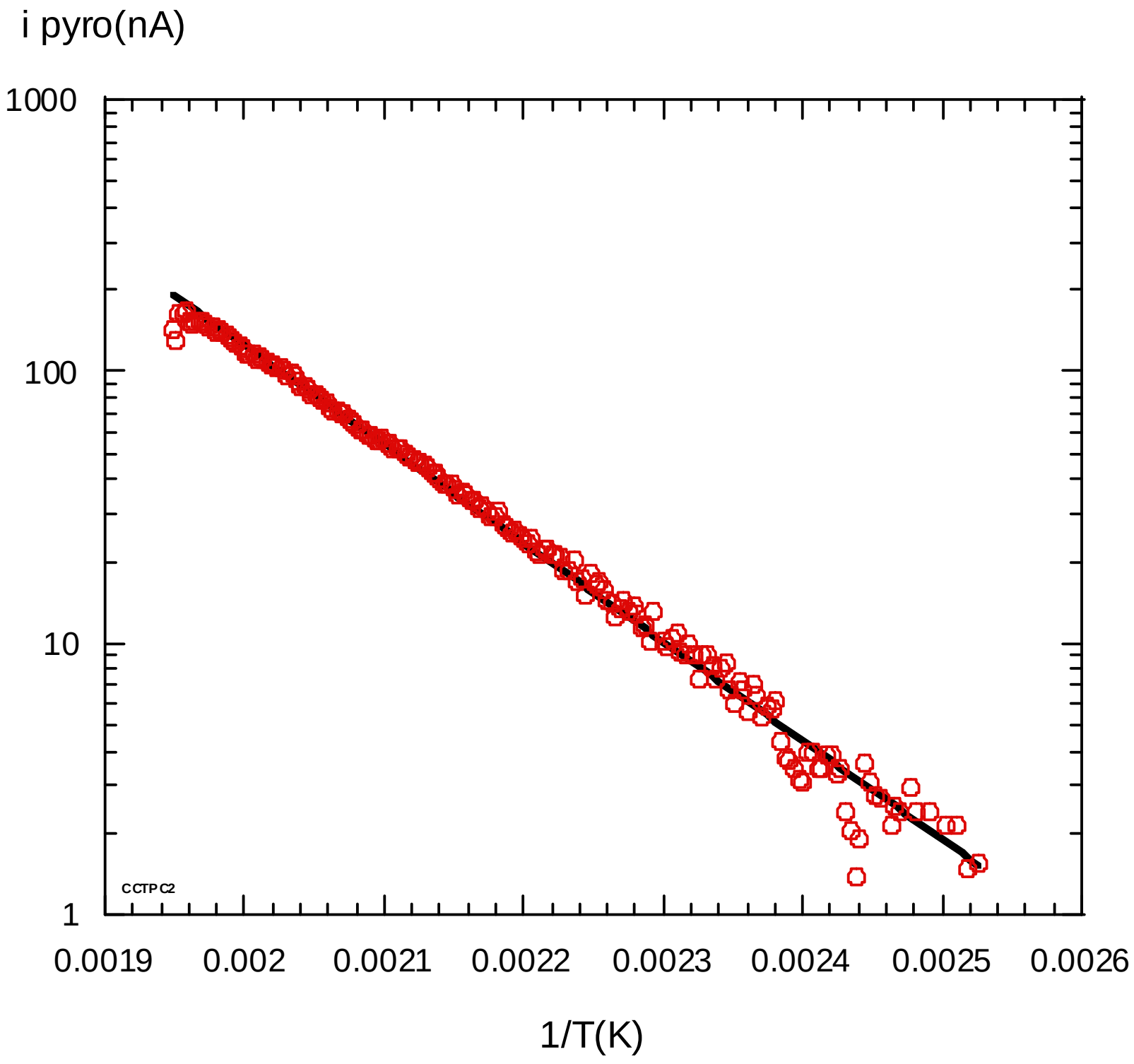

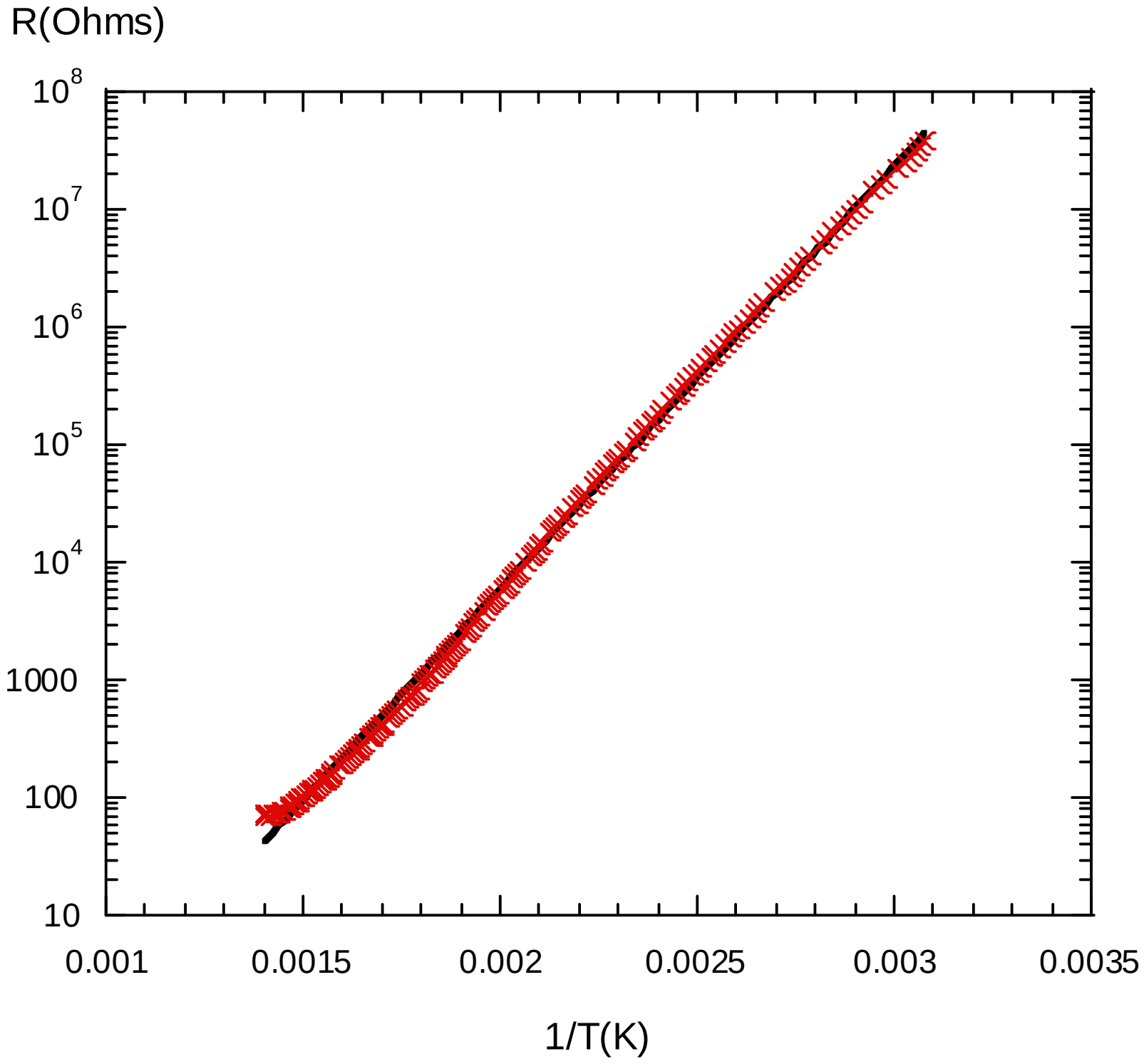

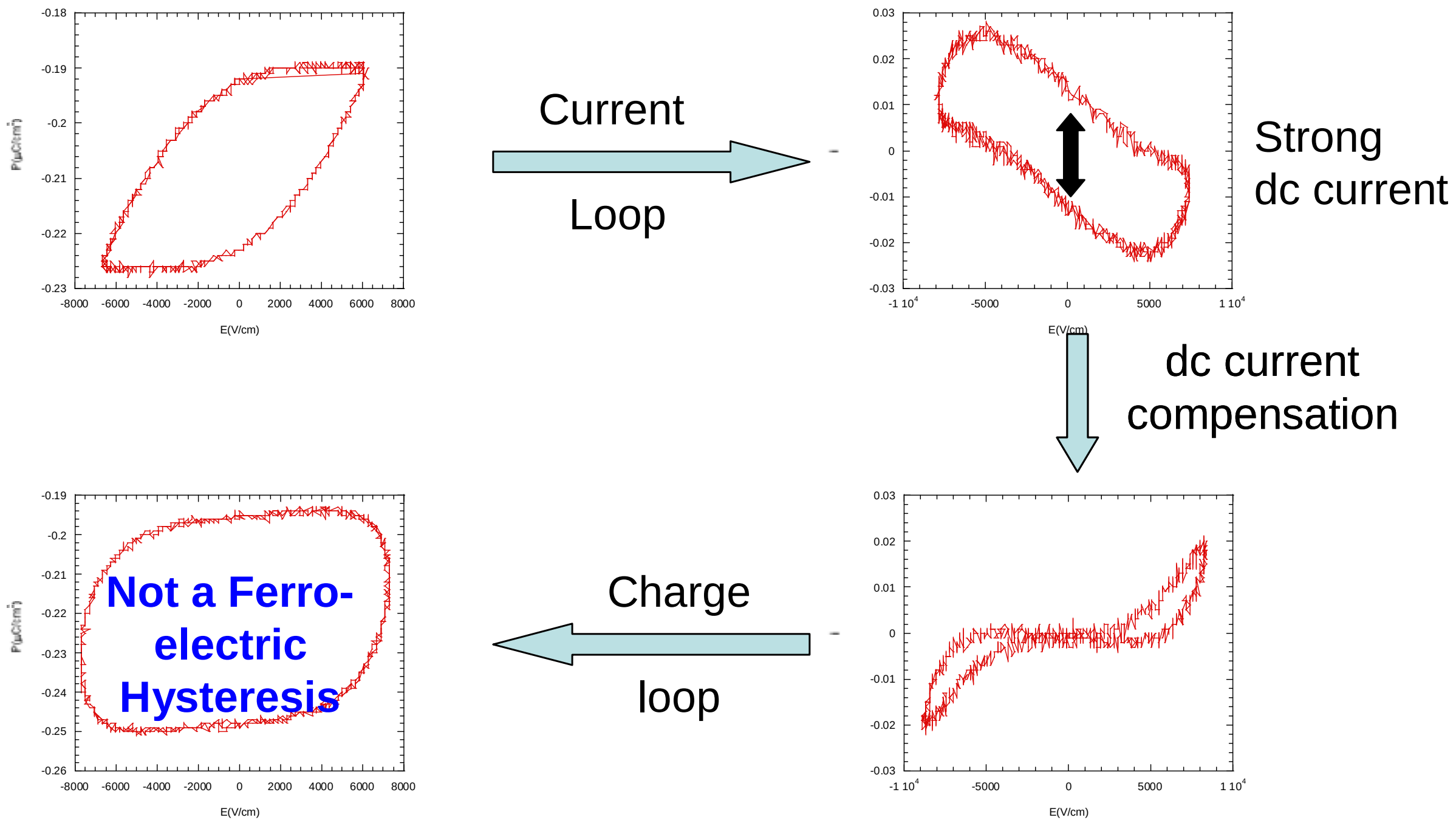

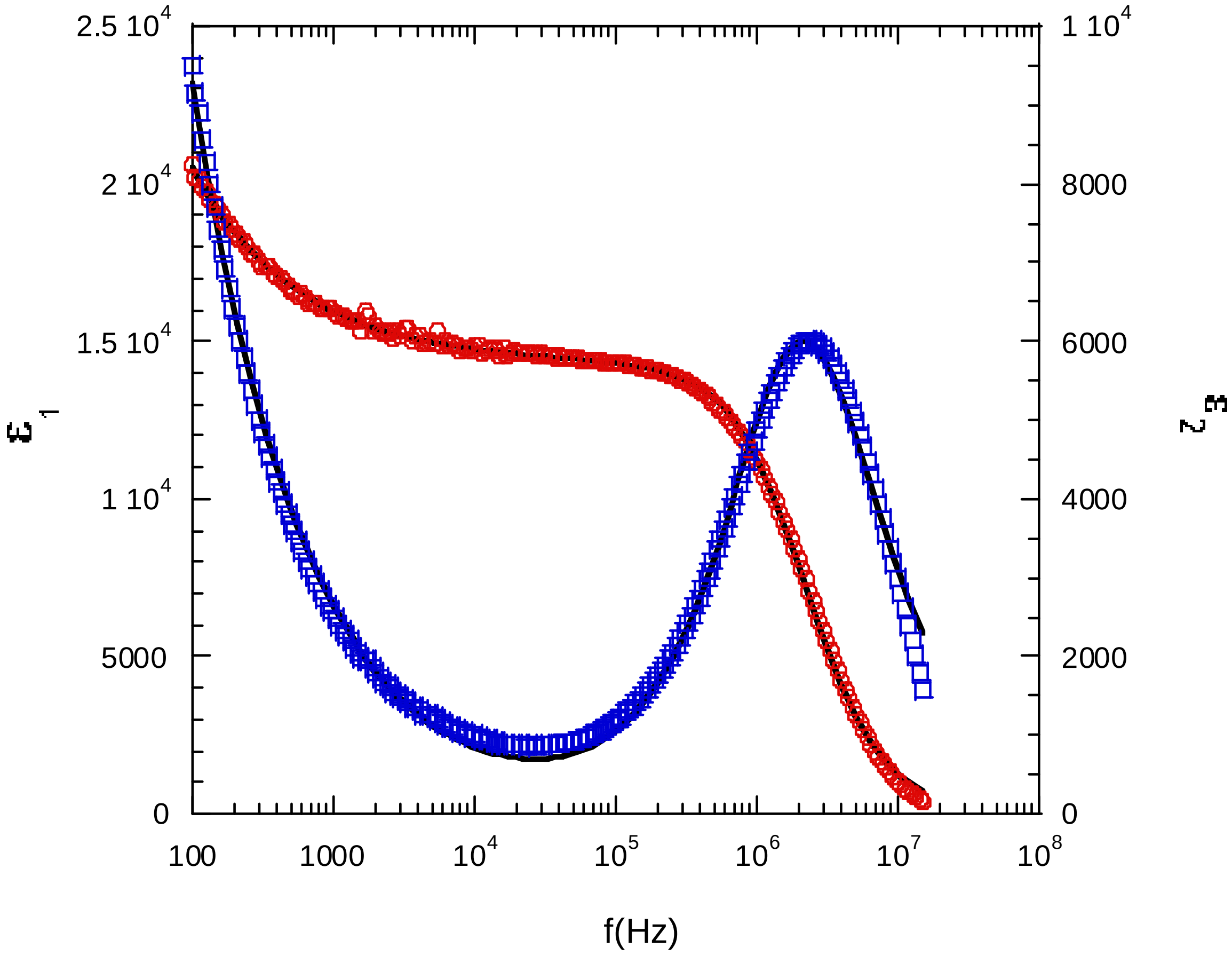